\documentclass[12pt,twoside]{article}
\usepackage[mathscr]{eucal}
\usepackage{amsmath,amsfonts,amssymb,amsthm}
\usepackage[matrix,arrow]{xy}
\usepackage{times}
\usepackage{pdfsync}
\usepackage{graphicx}% Include figure files
\usepackage{epstopdf}
\usepackage{dcolumn}
\usepackage{hyperref}
\usepackage{color}

\def\d_Vphi{\text{d}_V\hspace{-0.06em}\phi}
\def\d_Vphibar{\text{d}_V\hspace{-0.06em}\bar\phi}
\def\d_Vxi{\text{d}_V\hspace{-0.06em}\xi}

\def\be{\begin{eqnarray}}
\def\ee{\end{eqnarray}}
\def\beann{\begin{eqnarray*}}
\def\eeann{\end{eqnarray*}}
\def\beq{\begin{equation}}
\def\eeq{\end{equation}}
\def\ba{\begin{array}}
\def\ea{\end{array}}
\def\ben{\begin{enumerate}}
\def\een{\end{enumerate}}
\def\bea{\begin{eqnarray}}
\def\eea{\end{eqnarray}}

\def\5{\bar }
\def\6{\partial }
\def\7{\hat }
\def\4{\tilde }
\advance\textheight\topskip
\renewcommand{\tilde}{\widetilde}
\renewcommand{\hat}{\widehat}
\newtheorem{prop}{Proposition}[section]

\newtheorem{theorem}[prop]{Theorem}

\newcommand{\dd}{\partial}
\renewcommand{\d}{\partial}

\renewcommand{\geq}{\,{\geqslant}\,}
\renewcommand{\leq}{\,{\leqslant}\,}

\newcommand{\binner}[2]{%
  {\langle}\kern-4.15pt{\langle}#1{,}\,#2{\rangle}\kern-4.15pt{\rangle}}

\newcommand{\half}{\frac{1}{2}}

\newcommand{\ffrac}[2]{\raisebox{.5pt}%
  {\footnotesize$\displaystyle\frac{#1}{#2}$}\kern1pt}

\newcommand{\dover}[2]{\ffrac{\dd #1}{\dd #2}}

\def\cL{\mathcal{L}}

\def\cX{\mathcal{X}}
\def\cY{\mathcal{Y}}

\numberwithin{equation}{section} \makeatletter
\@addtoreset{equation}{section}

\DeclareFontFamily{OT1}{rsfs}{} \DeclareFontShape{OT1}{rsfs}{m}{n}{
<-7> rsfs5 <7-10> rsfs7 <10-> rsfs10}{}
\DeclareMathAlphabet{\mycal}{OT1}{rsfs}{m}{n}
\def\scri{{\mycal I}}%

\begin{document}

\def\mytitle{A note on the Newman-Unti group and the BMS charge
  algebra in terms of Newman-Penrose coefficients}

\pagestyle{myheadings} \markboth{\textsc{\small Barnich, Lambert}}{%
  \textsc{\small Newman-Unti group and BMS charge algebra}}
\addtolength{\headsep}{4pt}

\begin{flushright}\small
ULB-TH/11-01\end{flushright}

\begin{centering}

  \vspace{1cm}

  \textbf{\Large{\mytitle}}

%\vspace{1cm}

%{\huge Notes}

  \vspace{1.5cm}

  {\large Glenn Barnich$^{a}$ and Pierre-Henry Lambert$^{b}$}

\vspace{.5cm}

\begin{minipage}{.9\textwidth}\small \it \begin{center}
   Physique Th\'eorique et Math\'ematique\\ Universit\'e Libre de
   Bruxelles\\ and \\ International Solvay Institutes \\ Campus
   Plaine C.P. 231, B-1050 Bruxelles, Belgium \end{center}
\end{minipage}

\end{centering}

\vspace{1cm}

\begin{center}
  \begin{minipage}{.9\textwidth}
    \textsc{Abstract}. The symmetry algebra of asymptotically flat
    spacetimes at null infinity in four dimensions in the sense of
    Newman and Unti is revisited. As in the Bondi-Metzner-Sachs gauge,
    it is shown to be isomorphic to the direct sum of the abelian
    algebra of infinitesimal conformal rescalings with
    $\mathfrak{bms}_4$. The latter algebra is the semi-direct sum of
    infinitesimal supertranslations with the conformal Killing vectors
    of the Riemann sphere. Infinitesimal local conformal
    transformations can then consistently be included. We work out the
    local conformal properties of the relevant Newman-Penrose
    coefficients, construct the surface charges and derive their
    algebra.
  \end{minipage}
\end{center}

\vfill

\noindent
\mbox{}
\raisebox{-3\baselineskip}{%
  \parbox{\textwidth}{\mbox{}\hrulefill\\[-4pt]}}
{\scriptsize$^a$Research Director of the Fund for Scientific
  Research-FNRS Belgium. E-mail: gbarnich@ulb.ac.be\\$^b$
Boursier ULB. E-mail: Pierre-Henry.Lambert@ulb.ac.be}

\thispagestyle{empty}
\newpage

% \begin{small}
% {\addtolength{\parskip}{-1.5pt}
%  \tableofcontents}
% \end{small}
% \newpage

\section{Introduction}
\label{sec:introduction}

The definitions of asymptotically flat four dimensional
space-times at null infinity by Bondi-Van der Burg-Metzner-Sachs
\cite{Bondi:1962px,Sachs:1962wk} (BMS) and Newman-Unti (NU)
\cite{newman:891} in 1962 merely differ by the choice of
the radial coordinate. Such a change of gauge should not affect
the asymptotic symmetry algebra if, as we contend, this concept is
to have a major physical significance.

The problem of comparing the symmetry algebra in both cases is that,
besides the difference in gauge, the very definitions of these
algebras are not the same. Indeed, NU allow the leading part of the
metric induced on Scri to undergo a conformal rescaling. When this
generalization is considered in the BMS setting, it turns out that the
symmetry algebra is the direct sum of the BMS algebra
$\mathfrak{bms}_4$ \cite{Sachs2} with the abelian algebra of
infinitesimal conformal rescalings \cite{Barnich:2009se},
\cite{Barnich:2010eb}. There are two novel and independent aspects in
this computation.

\begin{itemize}

\item The first concerns the fact that the BMS algebra in $4$
  dimension involves the conformal Killing vectors of the unit, or
  equivalently, the Riemann sphere and can consistently accommodate
  infinitesimal local conformal transformations. The symmetry algebra
  $\mathfrak{bms}_4$ then involves two commuting copies of the non
  centrally extended Virasoro algebra, called superrotations in
  \cite{Barnich:2011ct}, and simultaneously the supertranslations
  generators are expanded in Laurent series. The standard, globally
  well-defined symmetry algebra $\mathfrak{bms}^{ \rm glob}_4$ consists in
  restricting to the globally well defined conformal Killing vectors
  of the sphere which correspond to infinitesimal Lorentz
  transformation, while the supertranslation generators are expanded
  into spherical harmonics.

  This local versus global versions of the symmetry algebra are of
  course not related to the BMS gauge choice, but will also occur in
  alternative characterizations of the asymptotic symmetry algebra
  where the conformal Killing vectors of the sphere play a
  role. Examples of this are the geometrical approach of Geroch
  \cite{Geroch:1977aa} based on Penrose's definition of null infinity
  \cite{PhysRevLett.10.66} and also, as we will explicitly discuss in
  this paper, the asymptotic symmetries in the NU framework.

\item The second aspect is related to the modified Lie bracket that
  should be used when the vector fields parametrising infinitesimal
  diffeomorphisms depend explicitly on the metric. Indeed, when using
  the modified Lie bracket, the space-time vectors realize the
  asymptotic symmetry algebra everywhere in the bulk and furthermore,
  even on Scri, this bracket is needed to disentangle the algebra when
  conformal rescalings of the induced metric on Scri are
  allowed. Similarly, in the context of the AdS/CFT correspondence,
  this bracket allows one to realize the asymptotic symmetry algebra
  in the bulk and to disentangle the symmetry algebra at infinity when
  considering transformations that leave the Fefferman-Graham ansatz
  invariant only up to conformal rescaling of the boundary metric
  \cite{Imbimbo:1999bj}. From a mathematical point of view, the
  modified Lie bracket is the natural bracket of the Lie algebroid
  that is associated to any theory with gauge invariance
  \cite{Barnich:2010xq}.

\end{itemize}

What we will do in this paper is to re-derive from scratch the
asymptotic symmetry algebra in the NU framework by focusing on metric
aspects and on the two novel features discussed above. As expected,
the symmetry algebra is again the direct sum of $\mathfrak{bms}_4$
with the abelian algebra of infinitesimal conformal rescalings of the
metric on Scri and thus coincides, as it should, with the generalized
symmetry algebra in the BMS approach. A related analysis of asymptotic
symmetries in the NU context from the point of view of Scri and
emphasizing global issues instead can be found in
\cite{0305-4470-11-1-012}, \cite{springerlink:10.1007/BF00669365}.

Even though the results presented here are not really surprising in
view of those in the BMS framework and the close relation between the
NU and BMS approaches, the exercise of working out the details is
justified because the NU framework is embedded in the context of the
widely used Newman-Penrose formalism \cite{newman:566} so that
explicit formulae in this context are directly relevant in many
applications, see e.g.~the  review article
\cite{newman:1980xx}.

As a first application, we study the transformation properties of the
Newman-Penrose coefficients parametrizing solution space in the NU
approach. Our main focus is on the inhomogeneous terms in the
transformation laws that contain the information on the central
extensions of the theory. We then discuss the associated surface
charges by following the analysis in the BMS gauge
\cite{Barnich:2011mi} and briefly compare with standard expressions
that can be found in the literature. The algebra of these charges is
derived and shown to involve field dependent central charges in
the case of $\mathfrak{bms}_4$ which vanish for $\mathfrak{bms}^{\rm
  glob}_4$.

\section{NU metric ansatz for asymptotically flat spacetimes}
\label{sec:from-bondi-van}

The metric ansatz of NU is based on a family of null hypersurfaces
labelled by the first coordinate, $x^0\equiv u={\rm const}$. The
second coordinate $x^1\equiv r$ is chosen as an affine parameter for
the null geodesic generators $l^\mu$ of these hypersurfaces, so that
$l^\mu=-\delta^\mu_r$. Up to a change of signature from $(+,-,-,-)$ to
$(-,+,+,+)$, a renumbering of the indices, and the tetrad
transformation that makes the conformal factor real, the line element
considered in section 4 of NU \cite{newman:891} can be written as
\begin{equation}
  \label{eq:10}
  ds^2=Wdu^2-2dr du+
g_{AB}(dx^A-V^Adu)(dx^B-V^Bdu)\,,
\end{equation}
with associated inverse metric
\begin{equation}
g^{\mu\nu}=  \begin{pmatrix} 0 &
  -1 & 0 \\
  -1&
-W  & -V^B \\
0 & -V^A & g^{AB}
\end{pmatrix}\,,\label{eq:NU}
\end{equation}
where
\begin{equation}
g_{AB}dx^Adx^B=r^2\bar\gamma_{AB}dx^Adx^B
+r C_{AB}dx^Adx^B+o(r)\,,\label{eq:11}
\end{equation}
with $\bar\gamma_{AB}$ conformally flat. Below, we will use standard
stereographic coordinates $\zeta=\cot{\frac{\theta}{2}}e^{i\phi},\bar
\zeta$, $\bar\gamma_{AB}dx^Adx^B=e^{2{\tilde\varphi}} d\zeta
d\bar\zeta$, ${\tilde\varphi}={\tilde\varphi}(u,x)$. In particular, we
use the notation $e^{2{\tilde\varphi}}$ for the conformal factor. In
section 4, we will give the explicit dictionary that allows one to
translate to the quantities originally used by NU.

In addition, the choice of origin for the affine parameter of the
null geodesics is fixed through the requirement that the
term proportional to $r^{-2}$ in the expansion of the spin
coefficient $-\rho=D_\rho l_\nu m^\rho\bar m^\nu$ is absent.

When expressed in terms of the metric, one finds
\begin{equation}
\rho=-\frac{1}{4} g^{AB}g_{AB,r}=-\frac{1}{4}\d_r
\ln |g| =-r^{-1}+\frac{1}{4}C^A_A r^{-2}+o(r^{-2})\,, \label{eq:36}
\end{equation}
where $g={\rm det}\, g_{\rho\nu}$ and the index has been raised with
the inverse of $\bar\gamma_{AB}$.  The requirement is thus equivalent
to the condition
\begin{equation}
  \label{eq:38}
  C^A_A=0\,.
\end{equation}
In the following we denote by $\bar D_A$ the covariant derivative with
respect to $\bar \gamma_{AB}$ and by $\bar\Delta$ the associated
Laplacian and by $\bar R$ the scalar curvature. In complex coordinates
$\zeta,\bar\zeta$, $C_{\zeta\bar\zeta}=0$ and we define for later
convenience $C_{\zeta\zeta}=e^{2{\tilde\varphi}} c,C_{\bar\zeta\bar
  \zeta}=e^{2{\tilde\varphi}} \bar c$. Finally,
\begin{equation}
  \label{eq:12}
V^A=O(r^{-2}),\qquad  W=-2 r\d_u
{\tilde\varphi}+\bar\Delta {\tilde\varphi}+O(r^{-1})\,,
\end{equation}
where $\bar \Delta
{\tilde\varphi}=4 e^{-2{\tilde\varphi}}\d\bar\d{\tilde\varphi}$ with $\d=
\d_\zeta,\bar\d=\d_{\bar\zeta}$.

The more restrictive fall-off conditions in \cite{newman:891} are
relevant for integrating the field equations but play no role in the
discussion of the asymptotic symmetry algebra.

\section{Asymptotic symmetries in the NU approach}
\label{sec:asympt-symm-newm}

The infinitesimal NU transformations can be defined as those
infinitesimal transformations that leave the form \eqref{eq:NU} and
the fall-off conditions \eqref{eq:11}-\eqref{eq:12} invariant, up to a
rescaling of the conformal factor
$\delta{\tilde\varphi}(u,x^A)={\tilde\omega}(u,x^A)$. In other words, they
satisfy
\begin{equation}
  \label{eq:4aext1}
  \cL_\xi g^{uu}=0,\quad \cL_\xi g^{uA}=0,\quad \cL_\xi
  g^{ur}=0,
\end{equation}
\begin{equation}
  \label{eq:39}
  \d_r \Big[\frac{1}{\sqrt{|g|}}\d_\rho (\sqrt {|g|} \xi^\rho)\Big]=o(r^{-2})\,,
\end{equation}
\begin{equation}
\begin{gathered}
\cL_\xi g^{rA}=O( r^{-2}),\quad  \cL_\xi
g^{AB}=-2 {\tilde\omega} g^{AB}+O( r^{-3}),\\
\cL_\xi g^{rr} =2r\d_u{\tilde\omega} +2{\tilde\omega} \bar\Delta
{\tilde\varphi} -\bar\Delta  {\tilde\omega} + O( r^{-1})\,.\label{eq:4aext2}
\end{gathered}
\end{equation}
Equations \eqref{eq:4aext1} are equivalent to
\begin{equation}
  \label{eq:4aextbis}
 \d_r \xi^\nu=g^{\nu\rho}\d_\rho \xi^u\iff \left\{
\begin{array}{l}\d_r\xi^u=0\,,\\ \d_r \xi^A=\d_B\xi^u g^{BA}\,,\\
 \d_r\xi^r=-\d_u\xi^u-\d_A \xi^u V^A\,,
\end{array}\right.
\end{equation}
and are explicitly solved by
\begin{gather}
  \label{eq:26}
\left\{\begin{array}{l}
  \xi^u=f,\\
\xi^A=Y^A+I^A, \quad  I^A=- \d_B f\int_r^\infty dr^\prime
g^{AB},\\
\xi^r=-r \d_u f+Z+ J,\quad J=\d_A f \int_r^\infty dr^\prime V^A,
\end{array}\right.
\end{gather}
with $\d_r f=0=\d_r Y^A=\d_r Z$. Equation \eqref{eq:39} then implies
\begin{equation}
  \label{eq:9}
  Z=\half \bar\Delta f\,.
\end{equation}
The first equation of \eqref{eq:4aext2} requires $\d_u Y^A=0$, the
second that $Y^A$ is a conformal Killing vector of $\bar\gamma_{AB}$,
which amounts to
\begin{equation}
  \label{eq:21}
  Y^\zeta \equiv Y=Y(\zeta),\quad  Y^{\bar \zeta} \equiv \bar Y=\bar
  Y(\bar \zeta)\,,
\end{equation}
in the coordinates $\zeta,\bar\zeta$, and also that
\begin{equation}
  \d_u f =f\d_u {\tilde\varphi}+\half \tilde \psi\,,
\label{eq:44ter}
\end{equation}
with $\psi=\bar D_A Y^A $, or more
explicitly in
$\zeta,\bar\zeta$ coordinates, $\psi= \d Y+\bar
\d\bar Y+2 Y \d{\tilde\varphi} +2\bar Y \bar \d {\tilde\varphi}$, and
$\tilde \psi=\psi-2{\tilde\omega}$.
Finally, the last equation of \eqref{eq:4aext2} implies
\begin{equation}
  \label{eq:24}
  2(\d_u Z+Z\d_u{\tilde\varphi})=Y^A\d_A
  \bar\Delta{\tilde\varphi}+\psi\bar \Delta{\tilde\varphi}+2 \d_A
  f\bar\gamma^{AB} \d_B\d_u{\tilde\varphi}+f\bar\Delta \d_u{\tilde\varphi}
-\bar\Delta {\tilde\omega},
\end{equation}
which is identically satisfied when taking the previous relations into
account.

One approach is to consider that \eqref{eq:44ter} fixes
${\tilde\omega}$ in terms of $f$ and $Y$,
${\tilde\omega}=\half\psi+f\d_u{\tilde\varphi}-\d_u f$.  Consider Scri, the space $\scri$
with coordinates $u,\zeta,\bar\zeta$ and metric
\begin{equation}
ds^2_{\scri}=0du^2+e^{2{\tilde\varphi}}d\zeta d\bar \zeta\,.\label{eq:16}
\end{equation}
The NU algebra is then defined as the commutator algebra of the vector
fields
\begin{equation}
\bar \xi =f\dover{}{u}+Y^A\dover{}{x^A}\,,\label{eq:17}
\end{equation}
with $f=f(u,x^A)$ arbitrary and $Y^A(x)$ conformal Killing vectors of
a conformally flat metric in $2$ dimensions, or equivalently, the
algebra of conformal vector fields of the degenerate metric
\eqref{eq:16}.

This is not the symmetry algebra of asymptotically flat spacetimes in
the sense of NU however. Indeed, ${\tilde\varphi}$ is arbitrary, it can for
instance be considered as the finite ambiguity related to Penrose's
conformal approach
\cite{PhysRevLett.10.66,penrose:1964,Penrose:1965am} to null
infinity. One can then interpret ${\tilde\varphi}$ as part of the background
structure, or in other words, of the gauge fixing
\cite{Geroch:1977aa}, and compute the asymptotic symmetries for a
fixed choice of ${\tilde\varphi}$, i.e., ${\tilde\omega}=0$ in the formulae above, or
ask the more general question of how the asymptotic symmetries depend
on changes in ${\tilde\varphi}$ by an arbitrary infinitesimal amount
${\tilde\omega}$. In both cases, one has to consider \eqref{eq:44ter} as a
differential equation for $f$. As we now show, the symmetry algebra
will then be isomorphic to the trivially extended $\mathfrak{bms}_4$
algebra by the abelian algebra of infinitesimal conformal rescalings,
as it should, and as a consequence, the Poincar\'e algebra is embedded
therein in a natural way. Furthermore, there is a natural realization
of the asymptotic symmetry algebra on an asymptotically flat 4
dimensional bulk spacetime. Note also that, for ${\tilde\omega}=0$, equation
\eqref{eq:44ter} has been interpreted from the point of view of
Penrose's conformal approach to null infinity in
\cite{0305-4470-11-1-012} following \cite{Tamburino:1966} and related
to the preservation of null angles, which is the standard way
\cite{PhysRevLett.10.66,penrose:1964,Penrose:1974,springerlink:10.1007/BF00762453}
to recover the BMS algebra from geometrical data on Scri.

The general solution for \eqref{eq:44ter} reads
\begin{equation}
  \label{eq:14}
  f=e^{
    {\tilde\varphi}}\big[ \tilde T+
  \half\int_0^u du^\prime
  e^{- {\tilde\varphi}}\tilde \psi\big],\quad  \tilde T= \tilde T(\zeta,\bar \zeta)\,,
\end{equation}
and the general solution to equations
\eqref{eq:4aext1}-\eqref{eq:4aext2} defining the asymptotic symmetries
is given by $\xi^\rho$ as in \eqref{eq:26} where $Z$,$Y^A$,$f$ satisfy
\eqref{eq:9}, \eqref{eq:21}, \eqref{eq:14} with ${\tilde\omega}$
arbitrary. Asymptotic Killing vectors thus depend on
$Y^A,\tilde T,{\tilde\omega}$ and the metric,
$\xi=\xi[Y,\tilde T,{\tilde\omega};g]$.

For such metric dependent vector fields, consider on the one hand the
suitably modified Lie bracket taking the metric dependence of the
spacetime vectors into account,
\begin{equation}
  \label{eq:43}
  [\xi_1,\xi_2]_M=[\xi_1,\xi_2]-\delta^g_{
    \xi_1}\xi_2+\delta^g_{ \xi_2}\xi_1,
\end{equation}
where $\delta^g_{\xi_1}\xi_2$ denotes the variation in $\xi_2$ under
the variation of the metric induced by $\xi_1$, $\delta^g_{
  \xi_1}g_{\mu\nu}=\cL_{\xi_1}g_{\mu\nu}$.

Consider on the other hand the extended $\mathfrak{bms}_4$ algebra,
i.e., the semi-direct sum of the algebra of conformal Killing vectors
of the Riemann sphere with the abelian ideal of infinitesimal
supertranslations, trivially extended by infinitesimal conformal
rescalings of the conformally flat degenerate metric on Scri. More
explicitly, the commutation relations are given by
$[(Y_1,\tilde T_1,{\tilde\omega}_1),(Y_2,\tilde T_2,{\tilde\omega}_2)]=(\hat Y,\hat
{\tilde T},\hat{{\tilde\omega}})$ where
\begin{equation}
  \left\{\begin{array}{l}
      \label{eq:5}\hat Y^A= Y^B_1\d_B
Y^A_2-Y^B_2\d_B Y^A_1,\\
\hat {\tilde T}=Y^A_1\d_A
\tilde  T_2-Y^A_2\d_A \tilde T_1 +\half (\tilde T_1\d_AY^A_2-\tilde T_2\d_AY^A_1),\\
\hat{{\tilde\omega}}=0\,.
\end{array}\right.
\end{equation}
It thus follows that
\begin{theorem}
  The spacetime vectors $\xi[Y,\tilde T,{\tilde\omega};g]$ realize the extended
  $\mathfrak{bms}_4$ algebra in the modified Lie bracket,
\begin{equation}
  \Big[\xi[Y_1,\tilde T_1,{\tilde\omega}_1;g],\xi[Y_2,\tilde T_2,{\tilde\omega}_2;g]\Big]_M=
\xi[\hat Y,\hat {\tilde T},\hat{{\tilde\omega}};g]\,,\label{eq:1}
\end{equation}
in the bulk of an asymptotically flat spacetime in the sense of Newman
and Unti.
\end{theorem}
Note in particular that for two different choices of the conformal
factor ${\tilde\varphi}$ which is held fixed, ${\tilde\omega}=0$, the
asymptotic symmetry algebras are isomorphic to $\mathfrak{bms}_4$,
which is thus a gauge invariant statement.

\proof{The proof follows closely the one in \cite{Barnich:2010eb} for
  the BMS gauge. In order to be self-contained we recall the different
  steps here.  In a first stage, one shows that on $\scri$, the
  vectors fields $\bar \xi[Y,\tilde T,{\tilde\omega};\bar\gamma]$ given in
  \eqref{eq:17} with $f$ as in \eqref{eq:14} realize the extended
  $\mathfrak{bms}_4$ algebra in terms of the modified Lie
  bracket. Indeed, this is obvious for the $A$ components which do not
  depend on the metric so that the modified bracket reduces to the
  standard Lie bracket for these components. For the $u$ component,
  taking into account that
\[\delta^g_{\bar \xi_1} f_2= {\tilde\omega}_1 f_2+\half e^{{\tilde\varphi}}\int_0^udu^\prime
e^{-{\tilde\varphi}}[-{\tilde\omega}_1(\psi_2-2{\tilde\omega}_2)+
2Y^A_2\d_A {\tilde\omega}_1]\,,\] we have
$[\bar\xi_1,\bar\xi_2]^u_M|_{u=0}= e^{\tilde\varphi}|_{u=0}\hat
T$. Direct computation then shows that
$\d_u([\bar\xi_1,\bar\xi_2]^u_M)=\hat f \d_u{\tilde\varphi}+\half\bar
D_A \hat Y^A$ with $\hat f$ given by \eqref{eq:14} with
$\tilde T,Y,{\tilde\omega}$ replaced by their hatted counterparts, implying
the result for the $u$ component.

For the spacetime vectors, direct computation gives
$[\xi_1,\xi_2]^u_M= [\bar\xi_1,\bar\xi_2]^u_M=\hat f$.  Using the
defining property \eqref{eq:4aextbis}, one then finds that
$\d_r([\xi_1,\xi_2]^\rho_M)=g^{\rho\nu}\d_\nu\hat f$. For the $A$
components the result then follows from the one on $\scri$,
$\lim_{r\to\infty} [\xi_1,\xi_2]_M^A=\hat Y^A$. This is due to the
fact that $I^A$ goes to zero at infinity, that the non-vanishing term
at infinity does not involve the metric and that the correction term
in the bracket does not change the asymptotic behaviour. Finally, for
the $r$ component, we still need to check that the $r$ independent
component of $[\xi_1,\xi_2]_M^r$ is given by $\half \bar\Delta \hat
f$, which follows by direct computation. \qed}

For completeness, let us also stress here that, if one focuses on
local properties and expands the conformal Killing vectors $Y^A\d_A$
and the infinitesimal supertranslations $T$ in Laurent series,
\begin{equation}
l_n=-\zeta^{n+1}\frac{\d}{\d\zeta},\quad \bar l_n=-\bar
\zeta^{n+1}\frac{\d}{\d\bar \zeta},\quad n\in \mathbb Z\,,\label{eq:55}
\end{equation}
\begin{equation}
\tilde  T_{m,n}=\zeta^m\bar\zeta^n, 
\quad m,n\in\mathbb Z\,, \label{eq:15}
\end{equation}
the commutation relations for the complexified
$\mathfrak{bms}_4$ algebra read
\begin{equation}
\begin{gathered}
  \label{eq:37}
  [l_m,l_n]=(m-n)l_{m+n},\quad [\bar l_m,\bar l_n]=(m-n)\bar
  l_{m+n},\quad [l_m,\bar l_n]=0, \\
[l_l,T_{m,n}]=(\frac{l+1}{2}-m)T_{m+l,n},
\quad [\bar l_l,T_{m,n}]= (\frac{l+1}{2}-n)T_{m,n+l}. 
\end{gathered}
\end{equation}
The $\mathfrak{bms}_4$ algebra contains as subalgebra the Poincar\'e
algebra, which we identify with the algebra of exact Killing vectors
of the Minkowski metric equipped with the standard Lie bracket.  It is
spanned by the generators
\begin{equation}
  l_{-1},\,l_0,\,l_1,\quad \bar l_{-1},\, \bar l_0,\, \bar l_1,\quad
\tilde T_{0,0},\,\tilde T_{1,0},\,\tilde T_{0,1},\,\tilde T_{1,1}\,.\label{eq:61}
\end{equation}

Non trivial central extensions of the algebra \eqref{eq:37} have been
studied in \cite{Barnich:2011ct}: the computation of
$H^2(\mathfrak{bms}_4)$ reveals that there are only the standard ones
for the Virasoro algebra extending the first two commutation
relations.

\section{Explicit relation between the NU and the BMS gauges}
\label{sec:expl-relat-with}
~
The definition of asymptotically flat space-times in the
BMS approach \cite{Bondi:1962px}, \cite{Sachs:1962wk},
\cite{Sachs2} as reviewed in \cite{Barnich:2009se},
\cite{Barnich:2010eb}, amounts to replacing $g_{uu}=1/g^{uu}=-1$ by
\begin{equation}
g_{uu}=1/g^{uu}=-e^{2\beta},\qquad \beta=O(r^{-2})\label{eq:6}
\end{equation}
in \eqref{eq:10} and \eqref{eq:NU} while imposing the additional
requirement that
\begin{equation}
{\rm det}\, g_{AB}=r^4{\rm det}\,\bar\gamma_{AB}\,.\label{eq:7}
\end{equation}

Both definitions then differ just by a choice of radial
coordinate. Indeed, replacing the radial coordinate by a function of
the $4$ coordinates preserves the zeros in \eqref{eq:10} and
\eqref{eq:NU} (see e.g.~the discussion in
\cite{0264-9381-20-19-302}). Furthermore, to first non trivial order
in $r$, the determinant condition leads to the same restriction
\eqref{eq:38} as the choice of the origin of the affine parameter. It
follows that the relation between the two radial coordinates does
not involve constant terms and is of the form
\begin{equation}
r^\prime=r +O(r^{-1})\label{eq:3}\,.
\end{equation}
More explicitly, starting from the NU approach,
BMS coordinates are obtained by defining the new
radial coordinates as \cite{Kroon:1999fk}
\begin{equation}
  \label{eq:2}
  r_{\rm BMS}=\big(\frac{{\rm det}\, g_{AB}}{{\rm det}\,\bar \gamma_{AB}}\big)^{\frac{1}{4}}\,.
\end{equation}
Conversely, starting from the BMS approach with radial coordinate $r$,
NU coordinates are obtained by changing the radial coordinate
to
\begin{equation}
  \label{eq:4}
   r_{\rm N} =r-\int^\infty_{ r} dr^\prime
  (e^{2\beta}-1)\,.
\end{equation}
These changes of coordinates only affect lower order terms in the
asymptotic expansion of the metric that play no role in the definition
of asymptotic symmetries and explains a posteriori why the asymptotic
symmetry algebras in both approaches are isomorphic.

At this stage, the dynamics of the theory comes into play. The
Einstein equations are solved order by order in $r$.  In the first
orders, there are integrations ``constants'' that appear as free data
characterizing asymptotically flat solutions. We will now work out the
explicit relation between these data in both approaches.  The inverse
metric in the BMS gauge (as discussed in \cite{Barnich:2010eb}) is
given by
\begin{equation}
g^{\mu\nu}_{BMS}=  \begin{pmatrix} 0 &
  -e^{-2\beta} & 0 \\
  -e^{-2\beta}&
-e^{-2\beta}\frac{V}{r}  & -e^{-2\beta}U^B \\
0 & -e^{-2\beta}U^A & g^{AB}
\end{pmatrix}\,.\label{eq:BMS}
\end{equation}
\begin{equation}
  \label{eq:96a}
  g_{AB}=r^2\bar\gamma_{AB}+rC_{AB}+\frac{1}{4} \bar\gamma_{AB}
C^C_DC^D_C+O(r^{-1}), 
\end{equation}
For simplicity, we assume here that there is no trace-free part
$D_{AB}$ at order $0$ and that the conformal factor is
time-independent, $\d_u\tilde\varphi=0$, in which case the news tensor
is simply $N_{AB}=\d_u C_{AB}$ and $f=T+\half u\tilde\psi$ with 
$T=e^{\tilde \varphi} \tilde T$. Writing
\begin{equation}
  C_{\zeta\zeta}=e^{2\tilde\varphi} c,\quad
  C_{\bar\zeta\bar\zeta}=e^{2\tilde\varphi} \bar c,\quad
  C_{\zeta\bar\zeta}=0,
\end{equation}
we have
\begin{equation}
\begin{split}
  \label{eq:30}
  \beta&=-\frac{1}{4}r^{-2}c\bar c +O(r^{-4}), \\
  U^\zeta&=-\frac{2}{r^{2}}e^{-4\tilde\varphi}\d(e^{2\tilde\varphi}\bar
  c)-\frac{2}{3r^3}\Big[ N^\zeta-4e^{-4\tilde\varphi}\bar c
\bar \d(e^{2\tilde\varphi} c)\Big]+O(r^{-4}),\\
  \frac{V}{r}&= 4e^{-2\tilde\varphi}\d\bar\d
  \tilde\varphi + r^{-1}2M +O(r^{-2}),
\end{split}
\end{equation}
which implies in particular that 
\begin{equation}
  \label{eq:13}
  r_{\rm N}=r+\frac{c\bar c}{2r}+O(r^{-3})\,.
\end{equation}
The only consequence of Einstein's equations on the angular momentum and mass aspects $N^\zeta=N^\zeta (u,\zeta,\bar\zeta),
M=M(u,\zeta,\bar\zeta)$ are the evolution equations 
\begin{equation}
  \label{eq:3a}
  \d_uM=-\frac{1}{8} N^A_BN^B_A
 +\frac{1}{8}
  \bar \Delta \bar R 
  +\frac{1}{4}\bar D_A\bar D_C N^{CA},
\end{equation}
\begin{multline}
  \label{eq:78ter}
\d_uN_A =\d_AM+\frac{1}{4}C_A^B\d_B\bar
  R +\frac{1}{16}\d_A\big[N^B_C C^C_B\big]
-\frac{1}{4} \bar D_AC^C_BN^B_C\\
-\frac{1}{4}\bar D_B\big[C^{B}_{C}N^C_{A}-N^B_CC^C_A\big]-\frac{1}{4}
  \bar D_B \big[ \bar D^B \bar D_CC^C_A -\bar D_A \bar
  D_CC^{BC}\big].
\end{multline}

Consider now the ``eth'' operators \cite{newman:863} defined
here for a field $\eta^s$ of spin weight $s$ according to the
conventions of \cite{Penrose:1986} through 
\begin{equation}
\eth \eta^s= P^{1-s}\bar \d(P^s\eta^s),\qquad \bar \eth
\eta^s=P^{1+s}\d(P^{-s}\eta^s)\,, \qquad P=\sqrt 2
e^{-\tilde\varphi}\,,
\label{eq:34}
\end{equation}
where
$\eth,\bar\eth$ raise respectively lower the spin weight by one
unit and satisfy  
\begin{equation} 
[\bar \eth, \eth]\eta^s=\frac{s}{2} \bar R\,
  \eta^s\,.\label{eq:35}
\end{equation}
The spin weights of the various quantities are summarized in 
table \ref{t1}. Note that the $P$ used here differs from the one used in
\cite{newman:891}, which we will denote by $P_N$ below. It also no
longer denotes the particular function $\half(1+\zeta\bar\zeta)$,
contrary to the notation used in
\cite{Barnich:2010eb,Barnich:2011mi}.

In order to compare with the notation used in \cite{newman:891}, we
use $\zeta=x^3+ix^4$. With $x^{\prime
  \alpha}=u,r_{\rm N},x^3,x^4$ and
$x^\mu=u,r,\zeta,\bar\zeta$,  computing $g^{\alpha\beta}_{\rm N}(x^\prime)=-\Big(\frac{\d
  x^{\prime\alpha}}{\d x^\mu}g^{\mu\nu}_{\rm BMS}\frac{\d
  x^{\prime\beta}}{\d x^\nu}\Big)(x(x^\prime))$, where the overall
minus sign takes the change of signature into account, then gives the
following dictionary by comparing with \cite{newman:891}:
\begin{equation}
  \begin{gathered}
    \label{eq:20}
    P_N=\frac{1}{\sqrt 2}e^{-{\tilde\varphi}}=\half P \,,\quad
    \nabla= 2\bar \d\,,
\quad \mu^0=-P^2\d\bar\d\ln P=\half \bar\Delta{\tilde\varphi}=-\frac{1}{4}\bar R\,,
    \\
    \Psi^0_{2}+\bar\Psi^0_{2}=-2M-\d_u{( c \bar c)}\,,\quad \sigma^0=
    \bar c\,,
\qquad \omega^0=\bar\eth \sigma^0\,,
    \\
    \Psi^0_{1}=-P N_{\bar\zeta}-\sigma^0\eth \bar\sigma^0-\frac{3}{4}
    \eth (\sigma^0\bar\sigma^0)\,.
  \end{gathered}
\end{equation}
For convenience, let us also use
\begin{equation}
  \label{eq:53}
  \Psi^0_{3}=-\eth
  \dot{\bar\sigma}^0-\frac{1}{4}\bar\eth\bar R, \qquad 
\Psi^0_{4}=-\ddot{\bar\sigma}^0\,.
\end{equation}
In these terms, 
\begin{equation}
  \label{eq:41}
\dot \Psi^0_3=\eth \Psi^0_4,\quad 
\dot \Psi^0_{2}=\eth
\Psi^0_{3}+ \sigma^0\Psi^0_{4},\qquad  \dot \Psi^0_{1}=\eth
\Psi^0_{2}+2 \sigma^0\Psi^0_{3}\,.
\end{equation}
Indeed, the first equation holds by definition and the assumed
time-independence of $P$. The evolution equation \eqref{eq:3a} is
equivalent to the real part of the second equation. Taking into
account the on-shell relation of the NU framework,
\begin{equation}
  \label{eq:48}
  \Psi^0_2-\bar\Psi^0_2=\bar\eth^2\sigma^0-\eth^2\bar\sigma^0
+\bar\sigma^0\dot \sigma^0-\sigma^0\dot{\bar\sigma}^0\,,
\end{equation}
we find 
\begin{equation}
  \label{eq:54}
M=-\Psi^0_{2}-\sigma^0\dot{\bar\sigma}^0+\half \bar\eth^2
\sigma^0-\half\eth^2\bar\sigma^0\,,
\end{equation}
in terms of which \eqref{eq:3a} is fully equivalent to the second
equation of \eqref{eq:41} and \eqref{eq:78ter} is equivalent to the
last equation of \eqref{eq:41}, in agreement with \cite{newman:891}.

\section{Transformation laws of the NU coefficients characterizing
  asymptotic solutions}
\label{sec:transf-laws-nu}

Let $\cY=P^{-1} \bar Y$ and $\bar \cY=P^{-1}
Y$. The conformal Killing equations and the conformal factor then
become
\begin{equation}
  \label{eq:25}
  \eth \bar \cY=0=\bar\eth \cY,\qquad \psi=(\eth \cY+\bar\eth
  \bar \cY)\,.
\end{equation}
It follows for instance that 
\begin{equation}
\bar\eth \eth \cY=-\frac{\bar
  R}{2}\cY,\quad \eth^2 \psi=\eth^3\cY-\frac{1}{2}\bar\cY\eth \bar R,\quad 
\bar\eth\eth \psi=-\frac{1}{2}[\eth(\bar R\cY)+\bar\eth(\bar
R\bar\cY)]\label{eq:44}\,.
\end{equation}
Using the notation $S=(Y,\tilde T,\tilde \omega)$, we have $-\delta_S
\bar \gamma_{AB}=2\tilde\omega \bar\gamma_{AB}$ for the background
metric and
\begin{equation}
  \label{eq:32}
  [-\delta_S,\bar \eth]\eta^s=-\tilde\omega\bar\eth\eta^s+s\bar\eth
  \tilde\omega\eta^s,\quad [-\delta_S, \eth]\eta^s=
-\tilde\omega\eth\eta^s-s\eth
  \tilde\omega\eta^s\,.
\end{equation}

To work out the transformation properties of the NU coefficients
characterizing asymptotic solution space, one needs to evaluate the
subleading terms in $\cL_\xi g^{\alpha\beta}_{N}$ on-shell. This can
also be done by translating the results from the BMS gauge, which
yields
\begin{equation}
\begin{split}
  \label{eq:16b}
  -\delta_S \sigma^0 & = [f\d_u+\cY\eth+ \bar
  \cY\bar\eth+\frac{3}{2}\eth \cY-\frac{1}{2} \bar\eth \bar \cY-\tilde
  \omega] \sigma^0-\eth^2
  f\,,\\
  -\delta_S \dot\sigma^0 & = [f\d_u+ \cY\eth + \bar \cY\bar\eth+2\eth
  \cY-2\tilde\omega]\dot\sigma^0-\half \eth^2\tilde \psi\,,\\
-\delta_S\Psi^0_4&=[f\d_u+\cY\eth+\bar\cY\bar\eth+\half \eth\cY
+\frac{5}{2}\bar\eth\bar\cY-3\tilde\omega]\Psi^0_4\,,\\
-\delta_S\Psi^0_3&=[f\d_u+\cY\eth+\bar\cY\bar\eth+\eth\cY
+2\bar\eth\bar\cY-3\tilde\omega]\Psi^0_3+\eth f\Psi^0_4\,,\\
  -\delta_S
  \Psi^0_2&=[f\d_u+\cY\eth+\bar\cY\bar\eth+\frac{3}{2}\eth \cY
  +\frac{3}{2}\bar\eth \bar\cY - 3\tilde\omega]\Psi^0_2
  +2\eth f\Psi^0_3,\\
  -\delta_S \Psi^0_1&
  =[f\d_u+\cY\eth+\bar\cY\bar\eth+2\eth\cY+\bar\eth\bar\cY
  -3\tilde\omega]\Psi^0_1  +3\eth f\Psi^0_2\,.
\end{split}
\end{equation}

Following for instance the terminology in \cite{held:3145} section 3,
but now for general infinitesimal transformations
$\zeta^\prime=\zeta+\epsilon Y(\zeta)$, $\bar \zeta^\prime=\bar
\zeta+\epsilon \bar Y(\bar \zeta)$ instead of those associated to
linear fractional transformations on the sphere and also considering
$\bar\zeta$ as the holomorphic coordinate instead of $\zeta$, a field
$\eta$ has spin weight $s$ and conformal weight $w$ if it transforms
as
\begin{equation}
  \label{eq:30a}
 - \delta_{Y,\bar Y} \eta=\big [Y^A\d_A+\frac{s}{2}(\bar \d
  \bar Y-\d Y)-\frac{w}{2}\psi\big] \eta\,.
\end{equation}
A tensor density of rank $s\geq 0$ and weight $n$ transforms as
\begin{equation}
  \label{eq:31a}
  -\delta_{Y,\bar Y} A_{\bar\zeta\dots\bar\zeta}=\big[Y^A\d_A + s \bar \d
  \bar Y+n(\d Y+\bar \d \bar Y) \big] A_{\bar\zeta\dots\bar\zeta}\,.
\end{equation}
while for rank $s\leq0$ and weight $n$, we have
\begin{equation}
  \label{eq:39s}
  -\delta_{Y,\bar Y} A_{\zeta\dots\zeta}=\big[Y^A\d_A - s  \d
  Y+n(\d Y+\bar \d \bar Y) \big] A_{\zeta\dots\zeta}\,.
\end{equation}
It then follows that a tensor density of weights $(s,n)$ defines a
field of weights $(s,-(2n+|s|))$ and conversely, a field of weights
$(s,w)$ defines a tensor density of weights $(s,-\half(w+|s|))$. For
$s\geq 0$, this is done through $\eta=
A_{\bar\zeta\dots\bar\zeta}P^{2n+s}$ and
$A_{\bar\zeta\dots\bar\zeta}=P^w\eta$. For $s\leq 0$, we have
$\eta= A_{\zeta\dots\zeta}P^{2n-s}$ and
$A_{\zeta\dots\zeta}=P^w\eta$. Note that complex
conjugation gives rise to opposite spin weight and rank but leaves
the conformal and density weights unchanged.
Alternatively, \eqref{eq:30a} can be written as 
\begin{equation}
  \label{eq:56}
  - \delta_{\cY,\bar \cY} \eta=\big [\cY\eth+\bar\cY\bar\eth +\frac{s-w}{2}\,\eth\cY
  -\frac{s+w}{2}\,\bar\eth\bar\cY\big] \eta\,.
\end{equation}

When focusing on $T=0=\tilde\omega$ at the surface $u=0$ and on the
homogeneous part of the transformations, this gives the weights
summarized in tables \ref{t1}, \ref{t2}. These tables are extended
to the Lie algebra elements, which are passive in all our
computations, by writing $[Y,\tilde T]=-\delta_{Y,\bar Y} \tilde T$
and $[Y,Y^\prime]^A=-\delta_{Y,\bar Y} Y^{\prime A}$.

\begin{table} 
\caption{Spin and conformal weights}\label{t1}
\begin{center}
\begin{tabular}{c|c|c|c|c|c|c|c|c} & $\sigma^0$  & $\dot\sigma^0$ &
  $\Psi^0_4 $&  $\Psi^0_3$ & $\Psi^0_2$ & $\Psi^0_1$ & $\cY$ & $T$  \\
\hline
s &  $2$ &  $2$  & $-2$  & $-1$ & $0$ & $1$ &  $-1$ & $0$ \\ 
\hline
w  & $- 1$  & $-2$  & $-3$ & $-3$ & $-3$ & $-3$ & $1$ & $1$ \\  
\end{tabular}
\end{center} \end{table}
\begin{table} 
\caption{Rank and density weights}\label{t2}
\begin{center}
\begin{tabular}{c|c|c|c|c|c|c|c|c}  & $P^{-1}\sigma^0$ &
  $P^{-2}\dot\sigma^0$ & $P^{-3}\Psi^0_4$ & 
$P^{-3}\Psi^0_3$ &  $P^{-3}\Psi^0_2$ & $
  P^{-3}\Psi^0_1$ & $\bar Y$ & $\tilde T$ \\
\hline
s  & $2$ & $2$ & $-2$ & $-1$ & 0 & $1$ &  $-1$  & $0$ \\ 
\hline
n  & $-\frac{1}{2}$ & 0 & $\half$ & $1$ & $\frac{3}{2}$ &
$1$ & $-1$ & $-\half$ \\  
\end{tabular}
\end{center} \end{table}

\section{Surface charge algebra}
\label{sec:surf-charge-algebra}

In this section, $\tilde\omega=0$ so that $f=T+\half u\psi$ and we use the
notation $s=(\cY,\bar\cY,T)$ for elements of the symmetry algebra,
which is given in these terms by $[s_1,s_2]=\hat s$ where 
\begin{equation}
\begin{gathered}
  \label{eq:57}
  \hat \cY=\cY_1\eth \cY_2 -(1\leftrightarrow 2),\qquad  \hat
  {\bar\cY}=\bar \cY_1\bar \eth \bar \cY_2 -(1\leftrightarrow 2),\\
  \hat T= (\cY_1\eth +\bar \cY_1\bar \eth)T_2-\half \psi_1 T_2 -(1\leftrightarrow 2)\,.
\end{gathered}
\end{equation}
The translation of the charges, the non-integrable piece due to the
news and the central charges computed in \cite{Barnich:2011mi} gives
here
\begin{flalign}
  Q_{s}[\cX]&=-\frac{1}{8\pi G}\int d^2\Omega^\varphi \Big[\big(
  f(\Psi^0_2+\sigma^0\dot{\bar
    \sigma}^0 )+\cY(\Psi^0_1
  +\sigma^0\eth\bar\sigma^0+\half\eth(\sigma^0\bar\sigma^0))\big)
  +{\rm c.c.}\Big],\nonumber\\
  \Theta_{s}[\delta\cX,\cX]&=\frac{1}{8 \pi G}\int d^2
  \Omega^\varphi\, f \big[\dot{\bar\sigma}^0\delta\sigma^0 +{\rm
    c.c.}\big]\,,  \label{eq:19}\\
  K_{s_1,s_2}[\cX]&= \frac{1}{8 \pi G}\int d^2 \Omega^\varphi\, \Big[
  \big(\frac{1}{4} f_1 \eth f_2 \bar\eth \bar R+\half \bar\sigma^0 f_1 \eth^2
  \psi_2 - (1\leftrightarrow 2) \big) + {\rm c.c.} \Big]\,.\nonumber
\end{flalign}
Note that one could also write the charges $Q_s[\cX]$ by allowing for
the additional terms $(\half \eth^2\bar\sigma^0-\half
\bar\eth^2\sigma^0)$ in the first parenthesis since these terms cancel
with the corresponding terms in the complex conjugate expression. Note
also that not $\Psi^0_2$ but only $\Psi^0_2+\bar\Psi^0_2$ is free data
on-shell because of the relation \eqref{eq:48}.

We recognize all the ingredients of the surface charges described in
\cite{0264-9381-1-1-005}, which in turn have been related there to
previous expressions in the literature and, in particular, to the
twistorial approach of Penrose \cite{Penrose08051982}. More precisely,
up to conventions, $Q_{0,0,T}$ agrees with Geroch's linear
super-momentum \cite{Geroch:1977aa} ${Q_{gn}}+\overline{Q}_{gn}$, as
given in equation (A1.12) of \cite{0264-9381-1-1-005}. The angular
(super-)momentum that we get is 
\begin{equation}
  Q_{\cY,0,0}=-\frac{1}{8\pi G}\int d^2\Omega^\varphi\,
  \cY\Big[\Psi^0_1
  +\sigma^0\eth\bar\sigma^0+\half\eth(\sigma^0\bar\sigma^0)
  - \frac{u}{2}\eth\big(\Psi^0_2+\bar\Psi^0_2+
  \d_u(\sigma^0\bar\sigma^0)\big)\Big]\,.
\label{eq:49}
\end{equation}
It differs from $Q_{\eta_c}$ given in equation (4) of
\cite{0264-9381-1-1-005} by the explicitly $u$-dependent term of the
second line.  It thus has a similar structure to Penrose's angular
momentum as described in equations (11), (12), and (17a) of
\cite{0264-9381-1-1-005} in the sense that it also differs by a
specific amount of linear supermomentum, but the amount is different
and explicitly $u$-dependent,
\begin{equation}
  \label{eq:50}
  Q_{\cY,0,0}= Q^{u=0}_{\cY,0,0}+\half u Q_{0,0,\eth\cY}\,.
\end{equation}

 The main result derived in \cite{Barnich:2011mi} states that 
\begin{itemize}

\item if one is allowed to integrate by parts, 
\begin{equation}
  \label{eq:27}
  \int d^2\Omega^\varphi\,\eth \eta^{-1} =0=\int
  d^2\Omega^\varphi\,\bar\eth \eta^{1},
\end{equation}
where $d^2\Omega^{\varphi}=\frac{2 d\zeta\wedge d\bar\zeta}{i P^2}$,

\item if one defines the ``Dirac bracket'' through
 \begin{equation}
   \label{eq:45}
   \{Q_{s_1},Q_{s_2}\}^*[\cX]=-\delta_{s_2}
   Q_{s_1}[\cX]+\Theta_{s_2}[-\delta_{s_1}\cX,\cX], 
 \end{equation}
\end{itemize}
then the charges define a representation of the $\mathfrak{bms}_4$
algebra, up to a field dependent central extension, 
\begin{equation}
  \label{eq:46}
  \{Q_{s_1},Q_{s_2}\}^*=Q_{[s_1,s_2]} +K_{s_1,s_2}, 
\end{equation}
where $K_{s_1,s_2}$ satisfies the generalized cocycle condition 
\begin{equation}
  \label{eq:47}
  K_{[s_1,s_2],s_3}-\delta_{s_3} K_{s_1,s_2}+{\rm cyclic} (1,2,3)=0\,.
\end{equation}
The representation theorem contained in equations \eqref{eq:46} and
\eqref{eq:47} can be verified directly in the present context by
starting from \eqref{eq:19}, \eqref{eq:48} and using the properties
\eqref{eq:35}, \eqref{eq:27} of $\eth$, the evolution equations
\eqref{eq:41}, the conformal Killing equations \eqref{eq:25}, the
$\mathfrak{bms}_4$ algebra \eqref{eq:57} and the transformation laws
\eqref{eq:16b}.

Several remarks are in order: 

\begin{itemize}

\item Integrations by parts are justified for regular functions on the
  sphere and thus for $\mathfrak{bms}^{\rm glob}_4$ and regular
  solutions. In the case of Laurent series more care is needed, see
  e.g.~\cite{saidi:1990xx}. We will address this question elsewhere.

\item For the globally well-defined $\mathfrak{bms}^{\rm glob}_4$
  algebra on the sphere, the central charge $K_{s_1,s_2}$ vanishes.

\item The non-conservation of the charges follows by taking
  $s_2=(0,0,1)$ and $s_1=s$. Indeed, since 
  $\frac{d}{du} Q_s=\frac{\d}{\d u} Q_s-\delta_{(0,0,1)} Q_s$, the
equality of the right hand sides of \eqref{eq:45}
 and \eqref{eq:46} gives 
 \begin{equation}
   \label{eq:51}
   \frac{d}{du} Q_s=-\frac{1}{8 \pi G}\int d^2
   \Omega^\varphi\, \big[\dot{\bar\sigma}^0(-\delta_s\sigma^0) +
   \frac{1}{4} \eth f\bar\eth \bar R+\half \bar\sigma^0 \eth^2
   \psi +{\rm
     c.c.}\big]
   \,.
 \end{equation}
 For $s=(0,0,1)$, this gives the standard Bondi-Sachs mass loss
 formula,
 \begin{equation}
   \label{eq:58}
   \frac{d}{du} Q_{0,0,1}= -\frac{1}{8 \pi G}\int d^2 \Omega^\varphi\,
 \big[\dot{\bar\sigma}^0\dot\sigma^0+{\rm c.c.}\big]\,.
 \end{equation}
 It also follows that the standard $\mathfrak{bms}^{\rm glob}_4$
 charges are all conserved on the sphere in the absence of news.
\end{itemize}

To the best of our knowledge, except for the previous analysis in the
BMS gauge, the above representation result does not exist elsewhere in
the literature.  A more detailed discussion of its implications, a
detailed comparison with results in the literature as well as a
self-contained derivation of the $\mathfrak{bms}_4$ transformation
laws in the context of the Newman-Penrose formalism will be given
elsewhere.

\section*{Acknowledgements}
\label{sec:acknowledgements}

\addcontentsline{toc}{section}{Acknowledgments}

The authors thank C\'edric Troessaert for useful discussions.  This
work is supported in part by the Fund for Scientific Research-FNRS
(Belgium), by the Belgian Federal Science Policy Office through the
Interuniversity Attraction Pole P6/11, by IISN-Belgium, by
``Communaut\'e fran\c caise de Belgique - Actions de Recherche
Concert\'ees'' and by Fondecyt Projects No.~1085322 and No.~1090753.

%\bibliography{/Users/gbarnich/Documents/Literature/Bibliography/master2}

\def\cprime{$'$}
\providecommand{\href}[2]{#2}\begingroup\raggedright\endgroup

\end{document}